\documentclass[sigconf,natbib=false]{acmart}
\AtBeginDocument{%
  }

\setcopyright{acmlicensed}

\copyrightyear{2027}
\acmYear{2027}
\acmDOI{XXXXXXX.XXXXXXX}
\acmConference[XXX]{XXXXXX}{April 5--9, 2027}{XXXXX}
\acmISBN{979-X-XXXX-XXXX-X/27/04}

\RequirePackage[
  datamodel=acmdatamodel,
  style=acmnumeric,
  ]{biblatex}

\begin{document}

\title{LLM Support for Mathematical Problem Posing: What Changes When Revision Formulations Are Supplied or Withheld?}
\renewcommand{\shorttitle}{LLM in Problem Posing}


\author{Shayla Sharmin}
\email{shayla@udel.edu}
\orcid{0000-0001-5137-1301}
\affiliation{%
  \institution{University of Delaware}
  \streetaddress{South College Avenue}
 \city{Newark}
  \state{Delaware}
  \country{USA}
 }
\author{Mohammad Fahim Abrar}

\email{fahim@udel.edu}
\orcid{0009-0009-5157-7807}
\affiliation{%
  \institution{University of Delaware}
  \streetaddress{South College Avenue}
 \city{Newark}
 \state{Delaware}
 \country{USA}
 }
\author{Mohammad Al-Ratrout}

\email{mratrout@udel.edu}
\orcid{0009-0009-5157-7807}
\affiliation{%
  \institution{University of Delaware}
  \streetaddress{South College Avenue}
 \city{Newark}
 \state{Delaware}
 \country{USA}
 }
\author{Aditya Raikwar}

\email{adirar@udel.edu}
\orcid{0009-0005-9427-9253}
\affiliation{%
  \institution{University of Delaware}
  \streetaddress{South College Avenue}
 \city{Newark}
 \state{Delaware}
 \country{USA}
 }
\author{Roghayeh Leila Barmaki}
\email{rlb@udel.edu}
\orcid{0000-0002-7570-5270}
\affiliation{%
  \institution{University of Delaware}
 \streetaddress{South College Avenue}
  \city{Newark}
  \state{Delaware}
  \country{USA}
 }
 
\renewcommand{\shortauthors}{anonymous et al.}

\begin{abstract}
In mathematical problem posing, participants construct or modify problems rather than only solve predefined ones, creating opportunities
for Large Language Models (LLMs) to support how those problems are formulated and revised. We investigate two forms of LLM support for revising participant-generated problems: Socratic scaffolding, which uses guiding questions while withholding candidate formulations, and Direct scaffolding, which provides concrete alternative formulations. We conducted a counterbalanced within-subjects study with 20 graduate
students who posed and revised problems across two mathematical scenarios. 
Problems showed larger immediate supported artifact-quality gains under Direct than Socratic scaffolding. Exploratory dimension-level analyses identified
statistically significant differences in structural clarity and novelty
after correction for multiple comparisons. Socratic scaffolding was
associated with longer revision time, highlighting a contrast between
immediate artifact outcomes and revision processes. These findings
identify the amount of candidate revision content supplied by an LLM
as an important design variable in mathematical problem posing.
\end{abstract}

 

\begin{teaserfigure}
\centering
\includegraphics[width=0.87\textwidth]
    {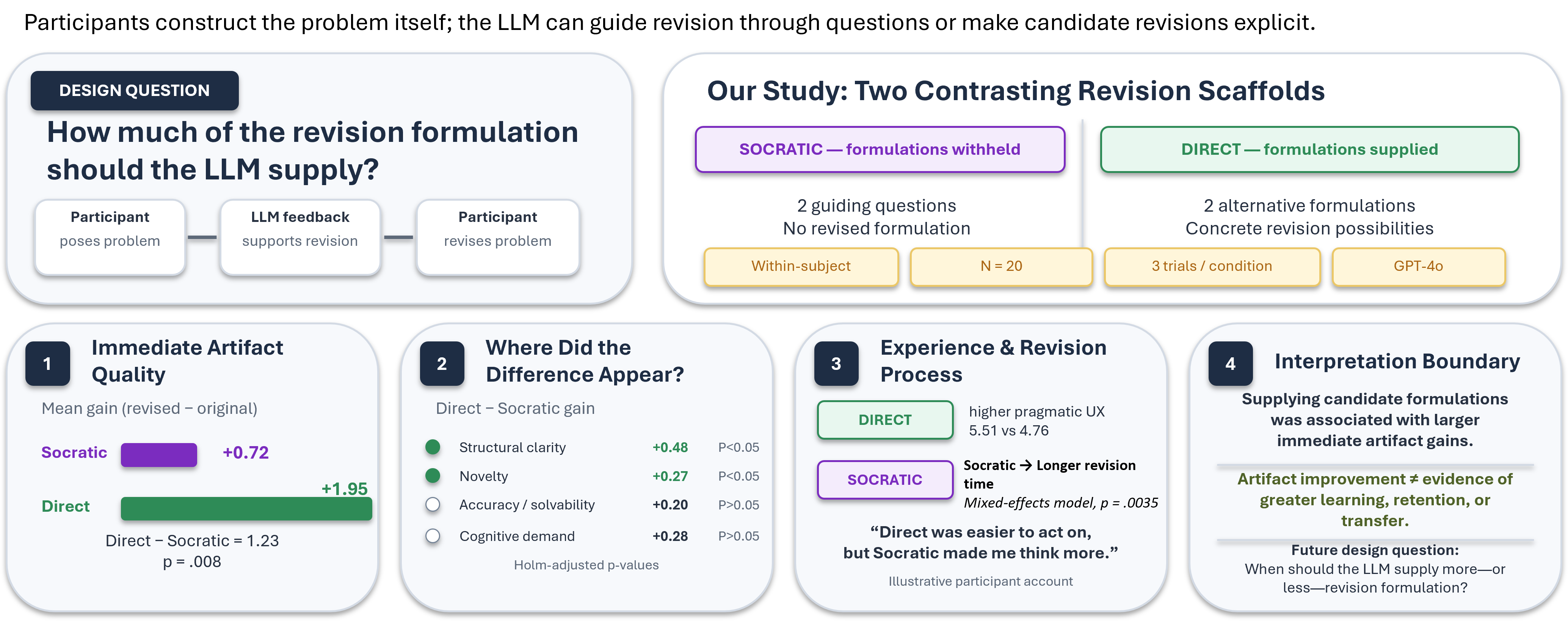}
\caption{Overview of the study design and main findings. We compare two
forms of LLM support for revising participant-generated mathematical
problems: Socratic scaffolding, which withholds candidate revision
formulations and provides guiding questions, and Direct scaffolding,
which supplies alternative formulations. Direct scaffolding was
associated with larger immediate artifact-quality gains, whereas
Socratic scaffolding involved longer revision time. Dimension-level
differences were clearest for structural clarity and novelty. These
artifact- and process-level differences do not establish corresponding
differences in learning.}
\Description{A study overview comparing Socratic and Direct LLM
scaffolding for mathematical problem posing. Participants first pose a
problem, receive LLM feedback, and revise the problem. Socratic
scaffolding withholds candidate formulations and provides two guiding
questions, whereas Direct scaffolding supplies two alternative
formulations. In a within-subject study with 20 participants and three
trials per condition, Direct scaffolding showed larger immediate
artifact-quality gains. The clearest adjusted dimension-level
differences were in structural clarity and novelty. Direct scaffolding
had higher pragmatic user-experience ratings, while revision time was
longer under Socratic scaffolding. An illustrative participant account
describes Direct feedback as easier to act on and Socratic feedback as
requiring more thought. The figure emphasizes that immediate artifact
improvement should not be interpreted as evidence of greater learning,
retention, or transfer.}
\label{fig:teaser}
\end{teaserfigure}

\received{}
\received[revised]{}
\received[accepted]{}

\maketitle
\section{Introduction}
\label{sec:introduction}

Problem posing is an established approach in mathematics education in which participants generate or modify mathematical problems rather than only solve problems provided to them~\cite{silver1994mathematical}. By constructing problems, participants decide what information and
relationships to include, what constraints to impose, and what reasoning the resulting problem should require. Problem posing has therefore been used to engage participants in mathematical thinking and to support conceptual understanding. Large Language Models (LLMs) have increasingly been used in educational activities to explain concepts, guide problem solving, provide tutoring support, and support programming activities~\cite{malik2025scaffolding}. 

Problem posing creates a different opportunity for LLM support because the participant is constructing the problem rather than working only toward a predefined answer. An LLM can therefore support not only the development of a posed problem but also its subsequent revision.
An LLM can support such revision in different ways. It can supply a
concrete candidate formulation that makes a possible revision explicit,
or it can provide guidance without supplying a revised formulation.
This distinction is particularly relevant to problem posing because the
object being revised is itself the participant-generated task. The LLM
can therefore guide revision while also potentially supplying content
that could appear in the revised problem.

In this study, we examine these forms of support as Direct and Socratic
scaffolding. Direct scaffolding provides concrete alternative
formulations that participants can consider during revision, whereas
Socratic scaffolding uses guiding questions without providing a revised
formulation. Related explicit and question-based forms of scaffolding
have been used in problem solving and other learning activities
~\cite{Brandmo2025Feedback,VENDRELL2026Scaffolding,
Krawitz2026ProblemPosing,ba2025investigating}. Recent work has explored generative AI in mathematical problems
~\cite{Walkington2025Middle,Kim2026AIM}.
We investigate how these two forms of LLM scaffolding support the
revision of participant-generated mathematical problems.
Socratic scaffolding guides revision through questions without
supplying candidate formulations, whereas Direct scaffolding
provides concrete alternatives. We examine the immediate
problem-quality outcomes, revision time, and user experience
associated with these designs to characterize their contrasting
patterns and inform how LLM revision support might be structured
for mathematical problem posing.~\autoref{fig:teaser} provides an overview of the study design, scaffolding contrast, and main findings.

In a counterbalanced within-subjects pilot study with 20 graduate
computing students, participants posed and revised problems across two
mathematical scenarios. 
We examine overall immediate quality change as the primary comparison
and explore condition differences across individual dimensions of problem
quality, with user experience and revision time reported as secondary
outcomes. Our research questions are:

\textbf{RQ1:} How do immediate changes in the quality of participant-generated
mathematical problems differ between Socratic and Direct LLM revision
scaffolding?


\textbf{RQ2:} How do changes in individual dimensions of problem quality
compare between Socratic and Direct LLM revision scaffolding?

Direct scaffolding was associated with a larger average immediate
quality gain than Socratic scaffolding. When the four quality criteria
were examined separately, significant differences after correction were
found for structural clarity and novelty, but not for conceptual accuracy
or cognitive demand. Together, these results show differences in the
immediate quality of problems revised with the two forms of LLM support.
This work contributes a controlled, within-participant comparison of
Socratic and Direct LLM scaffolding for revising participant-generated
mathematical problems. By comparing question-based guidance with concrete
alternative formulations, we examine how these contrasting forms of LLM
revision support relate to the immediate quality of the resulting
problems, both overall and across individual quality criteria. The
findings concern supported revision outcomes and do not establish
differences in learning.
\section{Related Work} \label{sec: related work}

%

\subsection{Problem Posing and LLM Support}

Problem posing asks participants to formulate, modify, or extend
problems rather than only solve problems defined for them
~\cite{silver1994mathematical,Silber2021ProblemPosing}. This distinction
has motivated the use of problem posing across educational settings,
including computer science, where it has been applied in introductory
programming as well as courses such as Data Structures and AI
~\cite{Mishra2015CSProblemPosing}. Because participants construct the
problem itself, the resulting artifact can vary in multiple properties,
including its validity, clarity, cognitive demand, and the extent to which
it introduces a meaningful variation. Support for problem posing must
therefore respond to an artifact whose content and structure are being
determined by the participant rather than only guide progress toward a
predefined solution.

Recent work has begun to examine how LLMs can participate in
problem-posing and related activities. In computing education, LLMs have
been explored in several roles. Students have posed problems and guided
an LLM through their solution process~\cite{yang2025learning}, while
question-based LLM support has been used to prompt reflection and clarify
goals during collaborative programming~\cite{Yan2025LLMcollaborative}.
LLMs have also been used to generate personalized programming exercises,
although prior work has identified challenges in areas such as difficulty
calibration and depth of personalization
~\cite{Logacheva2024ContextPersonalization}. These studies illustrate
that LLMs can participate in educational problem-related activities as
problem-solving partners, question-based supports, and generators of new
tasks.

Generative AI is also being investigated specifically in
mathematical problem posing. An LLM can question or prompt participants
as they develop a problem, but it can also generate candidate problem
content itself. Recent studies illustrate both the potential and the
difficulty of this capability: generative AI can expand the space of
problems available during problem-posing activities, while generated
problems may also contain mathematical or pedagogical limitations that
require evaluation~\cite{Walkington2025Middle, Kim2026AIM}. This work
shows that the role of an LLM in problem posing can extend beyond
providing general guidance to contributing content to the problem being
constructed.

However, generating or supporting the creation of a problem is
not the same as supporting the revision of a problem that a participant
has already posed. During revision, an LLM can identify aspects that
could be reconsidered while leaving the participant to determine how to
revise them, or it can make a possible revision explicit by supplying a
candidate formulation. This distinction matters because, in the latter
case, the support can contribute content directly to the artifact whose
quality is subsequently evaluated.

Our study focuses on this revision stage. Prior work establishes
broader uses of LLMs in problem posing, problem-related guidance, and
problem generation, but provides limited evidence comparing revision
support that supplies candidate formulations with support that withholds
them. We examine these two forms of support in the revision of
participant-generated mathematical problems and evaluate their association
with immediate changes in the quality of the resulting problems.
\subsection{The Assistance Dilemma in LLM Scaffolding}

A longstanding problem in instructional design is determining
when a system should provide information and when it should withhold
information so that participants must generate it themselves. Koedinger
and Aleven characterize this trade-off as the \emph{assistance dilemma}:
learning environments must balance assistance giving and assistance
withholding rather than assuming that either greater or lesser support is
uniformly preferable~\cite{KoedingerAleven2007Assistance}. Related scaffolding research similarly emphasizes contingency, fading, and the
gradual transfer of responsibility as central characteristics of
instructional support~\cite{vanDePol2010Scaffolding}. Our study does not implement a fading sequence; instead, it
examines one component of this broader design problem by holding the form
of support fixed within each condition and varying whether candidate
revision formulations are supplied.

LLMs make this dilemma especially salient because the information
they provide can include candidate content for the artifact being
constructed. This extends the assistance dilemma from support for completing a
task toward support that can contribute candidate content to the artifact
being evaluated, making the interpretation of immediate artifact quality
especially important. In mathematical problem posing, an LLM can guide revision
while withholding a candidate formulation, or it can make a possible
revision explicit by generating that formulation. Our Socratic and Direct
conditions operationalize this distinction at the level of immediate
problem revision. Thus, the relevant design variable is not simply a
choice between two feedback labels, but how much candidate revision
formulation the system supplies.

Prior instructional research also suggests that the appropriate
amount of explicit support may change over time. Worked-example and
fading research has shown benefits from initially making task-relevant
information explicit and subsequently increasing what learners must
generate themselves~\cite{RenklAtkinson2003Transition,
AtkinsonRenklMerrill2003Fading}. These findings arise from problem-solving
contexts rather than problem posing, so they do not predict which of our
conditions should produce better learning. Instead, they motivate treating
the amount of explicit LLM support as a design variable that may eventually
be adapted or faded rather than held fixed.

\subsection{Designing LLM Scaffolding for Revision}

LLMs make the explicitness of scaffolding unusually configurable:
the same system can ask questions, provide hints, explain a concept, or
generate content that can be incorporated directly into an artifact.
Accordingly, educational LLM research increasingly treats scaffolding as
a design problem rather than assuming that access to a capable model is
itself instructional support~\cite{Gong2026Asking, Faver2025Socratic,
Masalaci2025CustomGpt,malik2025scaffolding}. Pedagogical structures,
including question-based prompting and frameworks such as Bloom's
Taxonomy, have been used to constrain how LLMs provide support rather
than leaving interaction entirely open-ended.

At the same time, making assistance more explicit creates an
evaluation challenge. An LLM may help produce a stronger immediate output
without establishing that the participant has independently developed the
corresponding capability. Concerns about cognitive offloading and reliance
on generated content therefore motivate distinguishing supported artifact
performance from independent participant performance
~\cite{PEREZ2026Paradigm,VENDRELL2026Scaffolding}. This distinction is
particularly relevant to problem posing because the LLM can contribute
content to the very problem whose quality is subsequently evaluated.

Taken together, prior work motivates a design space rather than a
simple Direct-versus-Socratic ranking. LLM support can vary in how much
candidate revision content it makes explicit, while the consequences can
be evaluated separately for the resulting artifact and for participant
outcomes. We examine one controlled contrast within this space:
question-based Socratic support that withholds candidate formulations
versus Direct support that supplies alternative formulations. Our study
focuses on their association with immediate overall and dimension-level
changes in participant-generated mathematical problems.

A second distinction concerns what is being evaluated. Research
on learning and performance cautions that performance observed during
supported practice is not necessarily an indicator of durable learning:
conditions that improve immediate performance need not produce
corresponding gains in retention or transfer
~\cite{SoderstromBjork2015LearningPerformance}. This distinction is
particularly important when an LLM supplies candidate artifact content,
because the resulting artifact reflects performance produced with that
support available. We therefore treat immediate problem quality as an
artifact-level outcome and do not use it as evidence of independently
acquired problem-posing ability.
\section{Method}
\label{sec:method}

\subsection{Study Design}

We conducted a within-subjects, counterbalanced study to compare two
forms of LLM revision scaffolding---Socratic and Direct---in the revision
of participant-generated mathematical problems. Each participant
experienced both scaffolding conditions across two problem-posing
sessions. Each session paired one scaffolding condition with one
mathematical problem-posing scenario and contained three consecutive
problem-posing trials.

We counterbalanced condition and scenario order across four sequences, assigning five participants to each sequence. Thus, every
participant experienced both scaffolding conditions and both scenarios, but
experienced only one scaffolding condition within a given scenario.
Our Institutional Review Board (IRB) reviewed and approved the study protocol, and all participants provided informed consent before participation.

\subsection{Participants}

Twenty graduate students in computing participated voluntarily in the
study.
Participants were 23--34 years old (15 male, 5 female;
$M=29.7$).  Participation was uncompensated.
Participants reported substantial prior experience with LLMs, with median ratings of 7 for general LLM use and 6.5 for academic LLM use on 1--7 scales. Participants reported less experience with problem posing, with a median frequency rating of 4 for posing problems and 3 for being asked to create problems.

\subsection{Problem-Posing Tasks}

We adapted two mathematical problem-posing scenarios from prior research for this study~\cite{Silber2021ProblemPosing}. The scenarios involved mathematical content that was familiar to the
graduate participants. Accordingly, the task emphasized constructing
and revising problems rather than demonstrating advanced mathematical
knowledge. Although the underlying mathematics was elementary for this
sample, the posed problems could vary in clarity, solvability, cognitive
demand, and novelty.

\paragraph{\textbf{Handshake.}}
Participants were given the following situation:
``There are nine friends at a party. Each friend shakes hands with everyone
else when they arrive.'' The scenario supports graph- and combination-based reasoning.

\paragraph{\textbf{Making Change.}}
Participants were given the following situation:
``Sarah works at a coffee shop. A customer orders two small coffees at
\$1.59 each and pays with a \$5 bill. Sarah only has half-dollars, dimes,
and pennies.''
The scenario supports reasoning related to making change and coin
combinations.

During training, participants practiced posing a problem, requesting feedback, and revising their response using a separate scenario, \emph{Food Truck}, before beginning the experimental sessions to familiarize themselves with the interface and the main study. We answered their queries in this training session.

\subsection{LLM Scaffolding System}

Participants used a web-based problem-posing environment containing the
scenario, a text-entry area for constructing and revising a problem, and a feedback panel.  Once participants submitted their initial problems, the system generated feedback in real time using GPT-4o through the OpenAI API (temperature $=0.7$; maximum output $=350$ tokens).

Each trial separated problem construction, feedback inspection, and revision.
Participants first composed an initial problem and then reviewed the LLM feedback without being able to edit their response. Once the system re-enabled editing, they revised the same problem.
The complete prompts, condition-specific instructions, and additional study materials are available in the anonymous supplementary repository: \href{https://anonymous.4open.science/r/llm_problem_posing_supplementary_materials-73F0/}{[Anonymous Link]}.

\begin{figure*}[t]
    \centering
    \includegraphics[width=0.7\textwidth]{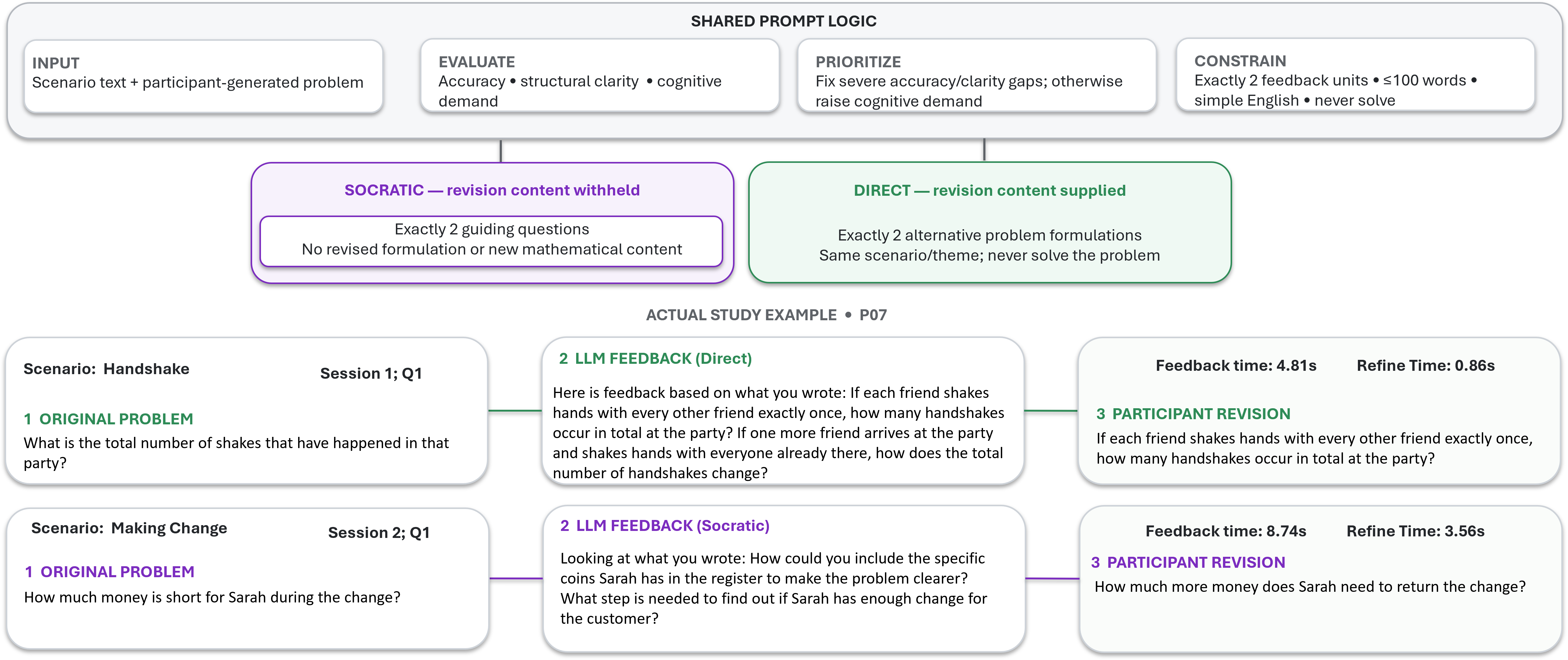}
    \caption{LLM scaffolding design and example study interactions.
Both conditions used the same input, evaluation criteria, feedback-priority
logic, and response constraints, but differed in whether feedback took
the form of guiding questions or concrete alternative formulations.
Socratic scaffolding returned two guiding questions without providing a
revised formulation, whereas Direct scaffolding returned two alternative
problem formulations. The lower panels show example interactions from one
participant, including the original problem, LLM feedback, and subsequent
participant revision. These examples illustrate how the two scaffolding
designs were instantiated in the study rather than serving as a comparative
case analysis.}
    \Description{Diagram showing the shared prompt logic and examples of
    Direct and Socratic LLM scaffolding. Both conditions receive a scenario
    and participant-generated problem, evaluate accuracy, structural clarity,
    and cognitive demand, apply the same feedback-priority rules, and return
    two feedback units without solving the problem. Direct scaffolding
    supplies two alternative problem formulations, while Socratic scaffolding
    asks two guiding questions without supplying revised content. Example
    study interactions show an original problem, the resulting LLM feedback,
    and the participant's revision for each condition.}
    \label{fig:scaffolding_example}
\end{figure*}

\subsection{Socratic and Direct Scaffolding}

Both scaffolding conditions used the same submitted problem,
evaluation criteria, feedback-priority rule, number of feedback units,
and response constraints. The conditions differed in how they presented feedback. Socratic scaffolding offered guiding questions without a revised formulation, whereas Direct scaffolding provided concrete alternatives.~\autoref{fig:scaffolding_example} illustrates the shared feedback-generation logic, the condition-specific scaffolding designs, and example interactions from the study.

In the \textbf{Socratic} condition, the LLM returned exactly two concise
guiding questions. These questions identified aspects of the problem that
could be reconsidered but were instructed not to provide a complete revised problem or prescribe the wording of the revision.

In the \textbf{Direct} condition, the LLM returned exactly two alternative
versions of the participant's problem. These alternatives supplied concrete revision possibilities while remaining within the information available in the assigned scenario.
Thus, Socratic scaffolding left the translation of feedback into
a revision unspecified, whereas Direct scaffolding made candidate
revisions explicit.

Before generating feedback, GPT-4o evaluated each submitted problem for conceptual accuracy, structural clarity, and cognitive demand. The prompt used Bloom's Taxonomy to define cognitive demand. Accuracy or clarity issues were prioritized when present;
otherwise, feedback targeted cognitive demand. The same evaluation logic was used in both conditions, which produced two feedback items in plain language (maximum 100 words) without solving the problem or introducing information outside the assigned scenario. We used the LLM's internal evaluation only to generate feedback, keeping it separate from the human-rated problem-quality scores used in our analysis.

\subsection{Procedure}

We organized the study into five phases: onboarding and training, the first problem-posing session, a post-session survey, the second problem-posing session with the same survey, and a 12--15 minute semi-structured interview.

\paragraph{\textbf{Onboarding and Training.}}
Participants completed informed consent and a demographic questionnaire and
received an introduction to mathematical problem posing and the study
interface. They were instructed to aim for four qualities when constructing
their problems: (1) accuracy, such that the problem had a correct answer;
(2) clarity, such that it was understandable to another person;
(3) cognitive challenge, requiring more than one reasoning step; and
(4) originality, rather than simply restating the scenario ~\cite{cai2013mathematical,cai2023impact,
Silber2021ProblemPosing}.

These instructional criteria corresponded conceptually to the four
dimensions later used to assess problem quality: conceptual accuracy and
solvability, structural clarity, cognitive demand, and novelty.
Participants then practiced the interaction procedure using the separate
\emph{Food Truck} scenario until they were comfortable with the interface.

\paragraph{\textbf{Experimental Sessions.}}
Each participant completed two experimental sessions. One session paired
one scenario with Socratic scaffolding and the other paired the remaining
scenario with Direct scaffolding. Condition and scenario order were
counterbalanced across the four study sequences described above.
Within each session, participants completed three consecutive trials. In each trial, they (1) read the assigned scenario, (2) composed an initial
problem, (3) requested LLM feedback, (4) reviewed the feedback, and
(5) revised and saved the same problem. During feedback presentation,
the problem-entry field was disabled and was re-enabled for revision.
Participants repeated this cycle three times per session.
After each experimental session, participants completed a post-session survey in Qualtrics.

Because participants received feedback across consecutive trials, their earlier feedback experiences may have influenced how they constructed later problems. Thus, our gain scores capture immediate changes during revision rather than the cumulative effects of feedback across trials. Although we counterbalanced condition order, we cannot rule out carryover effects between sessions.

\subsection{Measures}

\subsubsection{Problem Quality}

For each of the 120 trials, we retained the initially posed
problem and its revised version. We assessed problem quality using a four-dimension analytic rubric informed by prior mathematical problem-posing assessment literature~\cite{cai2013mathematical,
cai2023impact, Silber2021ProblemPosing}. The rubric assessed
\emph{conceptual accuracy and solvability}, \emph{structural clarity},
\emph{cognitive demand}, and \emph{novelty}, each scored from 0 to 2,
yielding a total quality score from 0 to 8 (\autoref{tab:rubric}).
A mathematics education expert reviewed the rubric for alignment
with the problem-posing task and scoring dimensions.

\begin{table}[t]
\centering
\footnotesize
\caption{Problem-quality rubric (0--2 per dimension).}
\label{tab:rubric}
\resizebox{\columnwidth}{!}{%
\begin{tabular}{lp{1.8cm}p{2cm}p{2.7cm}}
\toprule
\textbf{Dimension} & \textbf{Score 0} & \textbf{Score 1} & \textbf{Score 2} \\
\midrule
Conceptual Accuracy
& Incorrect or unsolvable
& Partially correct
& Correct and solvable \\

Structural Clarity
& Unclear or incomplete
& Some ambiguity
& Clear and complete \\

Cognitive Demand
& Remember/ Understand
& Apply
& Analyze/ Evaluate/ Create \\

Novelty
& Generic/ Restatement
& Some variation
& Clearly novel framing \\
\bottomrule
\end{tabular}%
}
\end{table}

One researcher rated all original and revised problems, and a second
researcher independently rated a randomly selected 20\% of the 240
problem artifacts ($n=48$) to assess inter-rater reliability. Both raters were blinded to the scaffolding condition during quality
assessment. Because the dimension scores were ordinal, reliability was
assessed using linearly weighted Gwet's AC2. Agreement was .86 for
conceptual accuracy and solvability, .70 for structural clarity, .89 for
cognitive demand, and .85 for novelty. The primary analyses used the
first rater's scores; the independently rated subset was used only to
assess inter-rater agreement.

\subsubsection{Manipulation Check and User Experience}

After each session, participants rated two manipulation-check items on a 7-point scale. The items asked whether the feedback (1) encouraged them to reflect on their own problem through questions rather than direct instructions and (2) directly showed them how to improve their problem. We also used the Short User Experience Questionnaire (UEQ-S) to assess pragmatic and hedonic user experience under each scaffolding condition as secondary, exploratory outcomes~\cite{hinderks2017design}. To assess subjective workload, participants rated mental demand, effort, and frustration on three NASA-TLX-derived items (0–100) after each condition. We averaged these three ratings to calculate an overall subjective-workload score for each condition~\cite{hart1988development}.

\subsubsection{Revision Time}

Revision time was recorded for each trial as the time spent revising the
problem after feedback. Because revision times were right-skewed and
repeated trials were nested within participants, we analyzed
log-transformed revision time using a linear mixed-effects model with
scaffolding condition and task position as fixed effects and a participant
random intercept. We treated revision time as a secondary behavioral measure rather than a direct measure of cognitive effort.

\subsubsection{Post-Study Interviews}

We used interview responses to provide context for the quantitative and behavioral findings, rather than conducting a standalone qualitative analysis. We selected quotations that helped explain participants' experiences related to the observed difference in revision time, without using them to estimate how common those experiences were.

\subsection{Statistical Analysis}

For each trial, quality gain was calculated as the revised total score
minus the original total score. Trial-level gains were then averaged
within participant and condition, yielding one mean gain for each
participant under Direct and Socratic scaffolding. These participant-level
mean gains constituted the primary outcome and were compared using a
paired $t$-test.
 A Wilcoxon signed-rank test served as a participant-level sensitivity
analysis. We also conducted trial-level sensitivity analyses using a
Gaussian GEE accounting for repeated observations within participants
and linear mixed-effects models with participant random intercepts.
Separate mixed-effects models additionally adjusted for trial position
and session order. We additionally examined a condition-by-scenario interaction as an
exploratory sensitivity analysis. Because each participant experienced
only one scaffolding condition within each scenario, this interaction
relied on the counterbalanced condition--scenario assignments rather than
a within-participant comparison of both conditions within the same
scenario.
Dimension-level comparisons were exploratory and Holm-adjusted across
the four rubric dimensions. Manipulation-check and user-experience
measures were compared using within-participant Wilcoxon signed-rank
tests and treated as secondary analyses. Revision time was analyzed
using the mixed-effects model described above. All tests were two-sided with $\alpha=.05$. Given the fixed sample size of 20 participants, a sensitivity analysis indicated that a two-sided paired-samples test (\(\alpha=.05\)) had 80\% power to detect a standardized within-participant effect of approximately \(d_z=0.66\) or larger.
\section{Results}
\label{sec:results}

\subsection{Problem Quality}

\subsubsection{Overall Quality Change}

Mean problem-quality scores were higher after revision than before
revision under both scaffolding conditions. Under Direct scaffolding, mean quality was $M=2.13$ before revision
and $M=4.08$ after revision, corresponding to a mean gain of
$1.95$ points ($SD=1.68$, 95\% CI $[1.17, 2.74]$). Under Socratic
scaffolding, the corresponding means were $M=2.00$ and $M=2.72$,
with a mean gain of $0.72$ points ($SD=0.80$, 95\% CI $[0.34, 1.09]$).

\begin{figure}[t]
    \centering
    \includegraphics[width=\columnwidth]{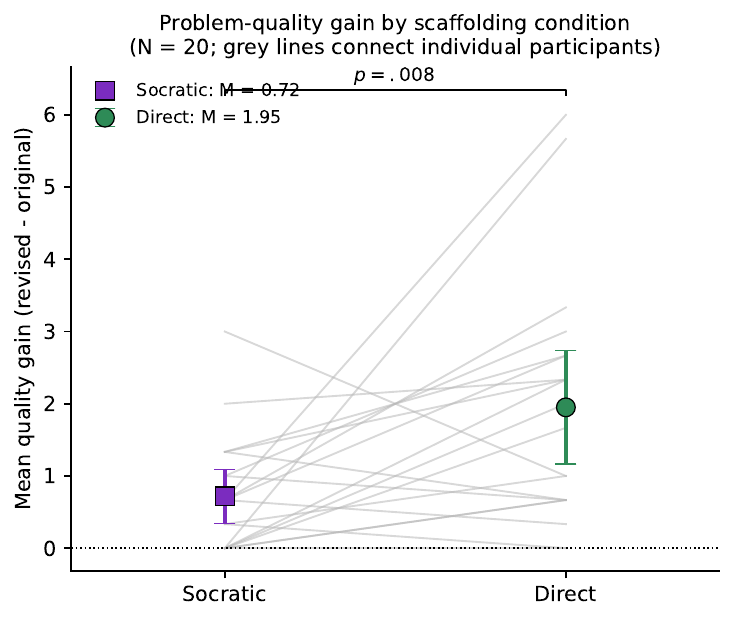}
    \caption{Participant-level mean problem-quality gains under Socratic
    and Direct scaffolding. Grey lines connect each participant's mean gain
    across conditions; colored markers show condition means with 95\%
    confidence intervals. The reported $p$-value corresponds to the primary
    paired comparison.}
    \Description{Paired plot of problem-quality gains for 20 participants.
    Each participant is connected across Socratic and Direct conditions.
    The mean gain is 0.72 under Socratic scaffolding and 1.95 under Direct
    scaffolding. Fourteen participants have larger gains under Direct,
    five have larger gains under Socratic, and one has equal gains.
    The paired comparison has p = .008.}
    \label{fig:paired_gain}
\end{figure}

The primary within-participant comparison showed that the mean quality gain
was $1.23$ points larger under Direct than Socratic scaffolding
(95\% CI $[0.36, 2.11]$), $t(19)=2.95$, $p=.008$,
Cohen's $d_z=0.66$. As shown in Figure~\ref{fig:paired_gain}, 14 of the
20 participants showed a larger mean gain under Direct scaffolding, five
showed a larger gain under Socratic scaffolding, and one showed equal gains.
A Wilcoxon signed-rank sensitivity analysis supported the same condition
difference ($W=25.5$, $p=.005$).

Trial-level sensitivity analyses accounting for repeated
observations supported the primary comparison, including a GEE model
(Direct--Socratic difference $=1.23$, 95\% CI $[0.43,2.03]$,
$p=.002$) and mixed-effects models adjusting for trial position and
session order (both $b=1.23$, $p<.001$).

\subsubsection{Quality Dimensions}

We next explored the Direct--Socratic difference separately for each
of the four dimensions of problem quality. Because these analyses were exploratory,
$p$-values were adjusted across the four comparisons using Holm's procedure.
The 95\% confidence intervals reported below are unadjusted and describe
the participant-level contrasts; statistical significance is evaluated
using the Holm-adjusted $p$-values.

\begin{figure}[t]
    \centering
    \includegraphics[width=\columnwidth]{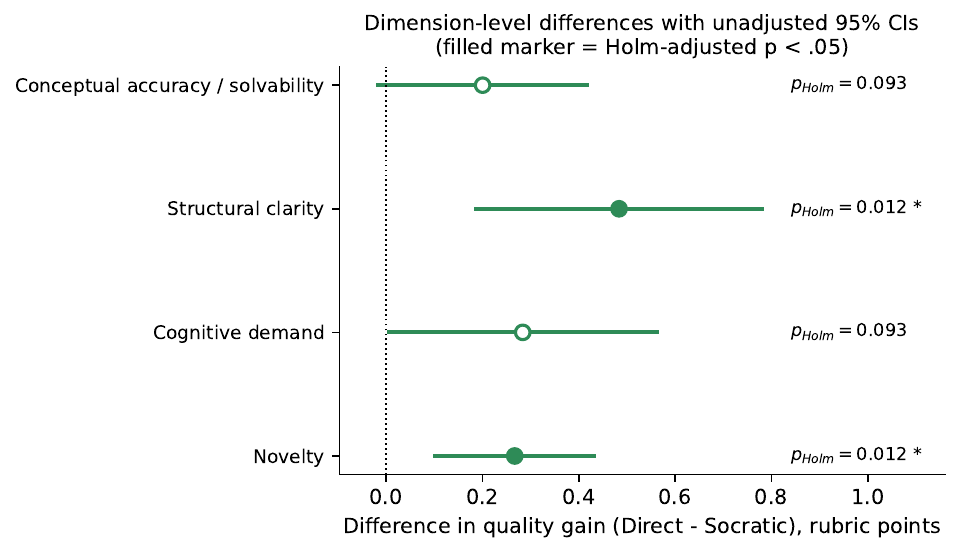}
    \caption{Direct--Socratic differences in problem-quality gain across
    the four rubric dimensions. Points show mean participant-level
    differences and horizontal lines show unadjusted 95\% confidence
    intervals. Filled markers indicate comparisons with
    Holm-adjusted $p<.05$; exact Holm-adjusted $p$-values are shown
    alongside each dimension.}
    \Description{Forest plot showing Direct minus Socratic differences in
    quality gain for conceptual accuracy and solvability, structural
    clarity, cognitive demand, and novelty. Structural clarity has a mean
    difference of 0.48 and novelty 0.27, both with Holm-adjusted p = .012.
    Conceptual accuracy and solvability has a difference of 0.20 and
    cognitive demand 0.28, both with Holm-adjusted p = .093.}
    \label{fig:dimensions}
\end{figure}

As shown in Figure~\ref{fig:dimensions}, Direct scaffolding showed a larger
gain than Socratic scaffolding in structural clarity
(mean difference $=0.48$, 95\% CI $[0.19, 0.78]$,
$p_{\mathrm{Holm}}=.012$) and novelty
(mean difference $=0.27$, 95\% CI $[0.10, 0.43]$,
$p_{\mathrm{Holm}}=.012$).

The differences in conceptual accuracy and solvability
(mean difference $=0.20$, 95\% CI $[-0.02, 0.42]$,
$p_{\mathrm{Holm}}=.093$) and cognitive demand
(mean difference $=0.28$, 95\% CI $[0.01, 0.56]$,
$p_{\mathrm{Holm}}=.093$) were not statistically significant after
Holm correction. As a sensitivity check, the Direct--Socratic difference was
descriptively larger for Making Change than Handshake, but the estimated
condition-by-scenario interaction was not statistically significant
(difference-in-differences $=1.27$, 95\% CI $[-0.40, 2.93]$,
$p=.127$).

\subsection{Manipulation Check}

Socratic system higher for prompting them with questions about their own
problem rather than telling them what to do
($M=5.05$, $SD=1.50$ vs.\ $M=3.80$, $SD=2.09$;
$W=18.0$, $p=.016$), whereas they rated the Direct system higher for
showing them exactly how to improve their problem
($M=5.30$, $SD=1.30$ vs.\ $M=4.35$, $SD=1.79$;
$W=22.5$, $p=.031$). Thus, participants' ratings showed relative differences
consistent with the intended manipulation.

\subsection{User Experience}

Direct scaffolding received higher pragmatic-quality ratings on the UEQ-S
($M=5.51$, $SD=0.76$ vs.\ $M=4.76$, $SD=0.94$;
$W=28.0$, $p=.004$) and higher overall UEQ-S scores
($M=5.25$, $SD=0.71$ vs.\ $M=4.69$, $SD=0.90$;
$W=26.5$, $p=.006$).

Hedonic-quality ratings did not significantly differ between Direct
($M=4.99$, $SD=1.02$) and Socratic scaffolding
($M=4.62$, $SD=1.15$; $W=41.5$, $p=.169$).
The three-item subjective workload from NASA-TLX score did not significantly differ between Direct ($M=42.28$, $SD=21.40$) and Socratic scaffolding ($M=49.80$, $SD=23.68$; $W=69.0$, $p=.189$).
These user-experience comparisons were secondary and exploratory; reported
$p$-values are unadjusted.

We additionally examined revision time as a secondary behavioral outcome.
After accounting for task position, the mixed-effects model estimated
revision time to be 2.03 times as long under Socratic as under Direct
scaffolding (95\% CI $[1.26, 3.26]$, $p=.0035$).

\section{Discussion}
\label{sec:discussion}

This study compared two forms of LLM support for revising
participant-generated mathematical problems: Socratic scaffolding,
which provided guiding questions without a revised formulation, and
Direct scaffolding, which provided concrete alternative formulations.
This comparison is particularly relevant to problem posing because the
LLM is responding to a problem constructed by the participant and can
support the revision of that problem in different ways. We found larger
immediate problem-quality gains under Direct than Socratic scaffolding.
Exploratory dimension-level analyses showed statistically significant
condition differences after correction for structural clarity and
novelty, but not for conceptual accuracy and solvability or cognitive
demand. Secondary results showed higher pragmatic and overall
user-experience ratings under Direct scaffolding, whereas revision
intervals were longer under Socratic scaffolding. Together, these
findings show that the two forms of LLM revision support were associated
with different immediate artifact and interaction outcomes.

\subsection{Immediate Quality Outcomes and Revision Experiences
Across Scaffolding Designs}

For RQ1, problems revised under Direct scaffolding showed larger
immediate quality gains than those revised under Socratic scaffolding.
One possible interpretation is that the concrete alternatives provided
under Direct scaffolding were more readily actionable during revision.
Participants could consider explicit formulations while deciding how
to revise their problems, whereas Socratic scaffolding offered guiding
questions without specifying a revised formulation. This interpretation
is consistent with the user-experience results, in which participants
rated Direct scaffolding higher in pragmatic quality and overall user
experience. Because Direct scaffolding supplied candidate formulations, its larger immediate quality gains may partly reflect participants’ incorporation or adaptation of LLM-generated content into their revisions. This possibility is particularly relevant for structural clarity and novelty, dimensions that could be directly influenced by the supplied formulations. Accordingly, the observed advantage should be interpreted as an advantage in supported artifact production rather than evidence that Direct scaffolding produced greater independent problem-posing ability.

The secondary behavioral results provide a complementary view of the
revision process. After accounting for task position, revision
intervals were approximately twice as long under Socratic scaffolding.
Illustrative interview accounts provide context for this finding.
P04 described Direct feedback as ``easier to act on'' while saying
that Socratic feedback ``made me think more,'' and P16 described
Socratic feedback as ``asking questions to make me mentally work,''
in contrast to Direct ``suggestions or specific ones.'' These accounts
illustrate how some participants experienced the two forms of support,
although longer revision intervals alone do not establish greater
cognitive effort.
Taken together, the findings reveal distinct patterns in immediate
artifact quality, revision time, and user experience. Direct
scaffolding was associated with larger immediate quality gains and
higher pragmatic ratings, whereas Socratic scaffolding was associated
with longer revision intervals. These differences highlight the
importance of examining both the resulting problem and the revision
process when evaluating LLM scaffolding for mathematical problem posing.

These findings also connect to the assistance dilemma, which concerns
the balance between providing information and leaving aspects of a
task for the participant to work out
~\cite{KoedingerAleven2007Assistance}. Direct scaffolding supplied
candidate revision formulations, whereas Socratic scaffolding guided
participants through questions without providing those formulations.
The observed differences illustrate how these two ways of structuring
assistance relate to immediate artifact and interaction outcomes.
They also motivate investigating when candidate formulations should
be supplied and when question-based guidance may be appropriate
during problem revision.

An important distinction remains between supported performance and
learning~\cite{SoderstromBjork2015LearningPerformance}. The present
findings characterize immediate revision outcomes under two forms of
LLM support, rather than changes in independent problem-posing
ability. Without an unassisted control condition or delayed post-test, we cannot determine whether either scaffolding approach produces greater learning, retention, or transfer to subsequent independent problem posing.
\subsection{Condition Differences Were Evident for Clarity and Novelty}

For RQ2, the exploratory analyses showed statistically significant
Direct--Socratic differences after Holm correction for structural clarity
and novelty, but not for conceptual accuracy and solvability or cognitive
demand. These results help unpack the overall quality difference by
identifying the dimensions with the clearest evidence of a condition
difference in this sample.
The structural-clarity result is consistent with the form of Direct
feedback. Providing complete alternative formulations gives participants
concrete examples of how a problem can be structured and expressed,
whereas Socratic feedback identifies aspects to reconsider through
questions without supplying a revised formulation. The larger clarity
gain under Direct scaffolding is therefore consistent with the greater
specificity of the support available during revision.

The novelty result is also notable. Direct feedback provided alternative
ways of formulating the participant's problem, potentially exposing
participants to problem framings that differed from their initial
versions. Some of this gain may therefore reflect the availability of candidate framings supplied by the LLM rather than participant-generated novelty alone. Problems revised under Direct scaffolding showed larger gains
on the rubric's novelty dimension. This finding concerns novelty as a
property of the resulting artifact rather than evidence of increased
independent creativity among participants.
The adjusted analyses did not establish condition differences for
conceptual accuracy and solvability or cognitive demand. These findings
highlight the value of examining individual dimensions rather than
relying solely on an aggregate quality score. However, because the
dimension-specific contrasts were not directly compared, statistical
significance for clarity and novelty should not be interpreted as
evidence of larger condition effects on these dimensions than on
accuracy or cognitive demand.

\subsection{Limitations and Future Directions}

This formative study involved 20 graduate computing students, two
mathematical scenarios, one LLM implementation, and two specific
scaffolding designs. The findings therefore need to be examined across
other populations, tasks, mathematical content, and LLM implementations. The sample size also limited statistical sensitivity: with 20 participants, the primary paired comparison had 80\% power only for effects of approximately \(d_z=0.66\) or larger. Smaller condition differences may therefore have gone undetected, particularly in the exploratory dimension-level analyses.

The mathematical content was relatively elementary for this graduate
sample, and future studies should examine whether similar patterns occur
when participants pose and revise problems involving more demanding or
less familiar content. The two conditions also represent particular
implementations of Socratic and Direct support; the findings should not
be generalized to all question-based or direct scaffolding designs.

Several aspects of the design also limit what can be concluded from the
observed changes. 
The study did not include an unassisted revision condition.
Although our design allowed us to compare immediate quality gains
between Direct and Socratic scaffolding, it could not isolate
improvement attributable to LLM support from improvement associated
with revision or practice alone. Repeated trials
also meant that later problems within a session were constructed after
participants had already received feedback under that condition, and
exposure to both conditions leaves open possible carryover despite
counterbalancing. In addition, feedback was generated dynamically by the
LLM, so individual feedback instances could vary within the same prompt
constraints.

Future work can address these limitations while examining how different
forms of support might be combined or adapted over time. Subsequent
unassisted tasks would help distinguish immediate supported artifact
quality from changes in independent problem-posing performance. Studies
with more demanding content and broader participant populations could
test the generality of the present findings. Scaffolding theory also
emphasizes contingent and fading support
~\cite{vanDePol2010Scaffolding}; accordingly, future studies could examine
adaptive or sequential designs in which question-based guidance and
concrete alternatives are introduced at different stages of revision
depending on participants' needs. 
Such work could extend the present findings by investigating
when question-based guidance, concrete revision formulations,
or combinations of both are most appropriate for different
problem-posing activities.

\section{Conclusion}
\label{sec:conclusion}

This study examined a central design choice in LLM-supported
mathematical problem posing: whether candidate revision formulations are
supplied or withheld during revision. Direct scaffolding, which supplied such formulations, was associated with larger immediate overall quality gains in the supported artifacts than Socratic scaffolding, although these differences were not uniform across dimensions of problem quality. The findings position the
amount of candidate revision content supplied by an LLM as an important
design variable rather than establishing either scaffolding approach as
generally preferable. They also underscore the nee  d to distinguish the
quality of an artifact produced with LLM support from learning or
independent problem-posing capability.

  \printbibliography

@article{hart1988development,
  title={Development of a multi-dimensional workload rating scale: Results of empirical and theoretical research},
  author={Hart, Sandra G},
  journal={Human mental workload},
  volume={1},
  pages={39--183},
  year={1988},
  publisher={Elsevier}, url={https://cir.nii.ac.jp/crid/1571698600629522048}
}

@article{hinderks2017design,
  title={Design and evaluation of a short version of the user experience questionnaire (UEQ-S)},
  author={Hinderks, Andreas and Thomaschewski, J{\"o}rg},
  journal={International Journal of Interactive Multimedia and Artificial Intelligence},
  year={2017}, url={https://www.academia.edu/download/54456371/ijimai20174_6_14_pdf_20309.pdf}
}

@article{Silber2021ProblemPosing,
  author = {Steven Silber and Jinfa Cai},
  title = {Exploring underprepared undergraduate students’ mathematical problem posing},
  journal = {ZDM -- Mathematics Education},
  year = {2021},
  volume = {53},
  number = {4},
  pages = {877--889},
  doi = {10.1007/s11858-021-01272-z},
  url = {https://doi.org/10.1007/s11858-021-01272-z}
}

@article{cai2013mathematical,
  author  = {Cai, Jinfa and Moyer, John C. and Wang, Ning and Hwang, Stephen and Nie, Bikai and Garber, Tammy},
  title   = {Mathematical Problem Posing as a Measure of Curricular Effect on Students' Learning},
  journal = {Educational Studies in Mathematics},
  volume  = {83},
  number  = {1},
  pages   = {57--69},
  year    = {2013},
  doi     = {10.1007/s10649-012-9429-3}
}

@article{cai2023impact,
  author  = {Cai, Jinfa and Ran, Hua and Hwang, Stephen and Ma, Yue and Han, Jaepil and Muirhead, Faith},
  title   = {Impact of Prompts on Students' Mathematical Problem Posing},
  journal = {The Journal of Mathematical Behavior},
  volume  = {72},
  pages   = {101087},
  year    = {2023},
  doi     = {10.1016/j.jmathb.2023.101087}
}

@article{Gong2026Asking,
  author  = {Gong, Yulin and Wang, Minkai and He, Li and Xu, Chengshu and Yu, Yue},
  title   = {Asking, Playing, Learning: Investigating Large Language Model-Based Scaffolding in Digital Game-Based Learning for Elementary Artificial Intelligence Education},
  journal = {Journal of Educational Computing Research},
  volume  = {64},
  number  = {2},
  pages   = {311--343},
  year    = {2026},
  doi      = {10.1177/07356331251396354},
  url      = {https://doi.org/10.1177/07356331251396354}
}

@article{Walkington2025Middle,
  author  = {Walkington, Candace and Pando, Magdalena and Lipsmeyer, Lin Lin and Beauchamp, Theodora and Sager, Marc and Milton, Saki},
  title   = {Middle School Girls Using Generative AI to Engage in Mathematical Problem-Posing},
  journal = {Mathematical Thinking and Learning},
  pages    = {1--22},
  year     = {2025},
  doi      = {10.1080/10986065.2025.2542724},
  url      = {https://doi.org/10.1080/10986065.2025.2542724}
}

@article{VENDRELL2026Scaffolding,
title = {Scaffolding critical thinking with generative AI: Design principles for integrating large language models in higher education},
journal = {Computers and Education: Artificial Intelligence},
volume = {10},
pages = {100572},
year = {2026},
issn = {2666-920X},
doi = {https://doi.org/10.1016/j.caeai.2026.100572},
url = {https://www.sciencedirect.com/science/article/pii/S2666920X26000342},
author = {Mireia Vendrell and Samantha-Kaye Johnston}
}

@inproceedings{Masalaci2025CustomGpt,
author = {Masalaci, Zeynep and Shalawadi, Sujay and Papachristos, Eleftherios},
title = {From ChatGPT to Custom GPTs: Scaffolding Conceptual Learning in Adult Education},
year = {2025},
isbn = {9798400720154},
publisher = {Association for Computing Machinery},
address = {New York, NY, USA},
url = {https://doi.org/10.1145/3771882.3771907},
doi = {10.1145/3771882.3771907},
booktitle = {Proceedings of the 24th International Conference on Mobile and Ubiquitous Multimedia},
pages = {390–401},
numpages = {12},
location = {
},
series = {MUM '25}
}

@article{PEREZ2026Paradigm,
title = {Large language models for education: An open-source paradigm for automated Q\&A in the graduate classroom},
journal = {Computers and Education: Artificial Intelligence},
volume = {10},
pages = {100546},
year = {2026},
issn = {2666-920X},
doi = {https://doi.org/10.1016/j.caeai.2026.100546},
url = {https://www.sciencedirect.com/science/article/pii/S2666920X2600007X},
author = {Ryann M. Perez and Marie Shimogawa and Yanan Chang and Xinning Li and Hoang Anh T. Phan and Jason G. Marmorstein and Evan S.K. Yanagawa and E. James Petersson},
}

@article{malik2025scaffolding,
  title={Scaffolding middle school mathematics curricula with large language models},
  author={Malik, Rizwaan and Abdi, Dorna and Wang, Rose and Demszky, Dorottya},
  journal={British Journal of Educational Technology},
  volume={56},
  number={3},
  pages={999--1027},
  year={2025},
  publisher={Wiley Online Library},
  doi={https://doi.org/10.1111/bjet.13571}
}

@article{silver1994mathematical,
  title={On mathematical problem posing},
  author={Silver, Edward A},
  journal={For the learning of mathematics},
  volume={14},
  number={1},
  pages={19--28},
  year={1994},
  publisher={JSTOR},
  url={http://www.jstor.org/stable/40248099}
}

@ARTICLE{Brandmo2025Feedback,
    
AUTHOR={Brandmo, Christian  and Gamlem, Siv M. },
           
TITLE={Students' perceptions and outcome of teacher feedback: a systematic review},
          
JOURNAL={Frontiers in Education},
          
VOLUME={Volume 10 - 2025},
  
YEAR={2025},
  
URL={https://www.frontiersin.org/journals/education/articles/10.3389/feduc.2025.1572950},
  
DOI={10.3389/feduc.2025.1572950},
  
ISSN={2504-284X}
}

@InProceedings{Faver2025Socratic,
author={Favero, Lucile
and P{\'e}rez-Ortiz, Juan Antonio
and K{\"a}ser, Tanja
and Oliver, Nuria},
editor={Bellas, Francisco
and Fontenla-Romero, Oscar},
title={Enhancing Critical Thinking in Education by Means of a Socratic Chatbot},
booktitle={AI in Education and Educational Research},
year={2025},
publisher={Springer Nature Switzerland},
address={Cham},
pages={17--32},
isbn={978-3-031-93409-4},
doi={ https://doi.org/10.1007/978-3-031-93409-4_2}
}

@article{Krawitz2026ProblemPosing,
  author  = {Krawitz, Jan and Meyer-Jen{\ss}en, Lisa and Krausm{\"u}ller, Katharina and others},
  title   = {Problem posing and motivation: the effects of posing and solving one’s own modelling problems on autonomy, competence, relatedness, and self-efficacy},
  journal = {ZDM Mathematics Education},
  year    = {2026},
  doi     = {10.1007/s11858-025-01762-4},
  url     = {https://doi.org/10.1007/s11858-025-01762-4}
}

@article{ba2025investigating,
  title={Investigating the impact of ChatGPT-assisted feedback on the dynamics and outcomes of online inquiry-based discussion},
  author={Ba, Shen and Zhan, Ying and Huang, Lingyun and Lu, Guoqing},
  journal={British Journal of Educational Technology},
  volume={56},
  number={5},
  pages={1710--1734},
  publisher={Wiley Online Library},
  doi={https://doi.org/10.1111/bjet.13605}
}

@article{Mishra2015CSProblemPosing,
  author  = {Mishra, Shitanshu and Iyer, Sridhar},
  title   = {An exploration of problem posing-based activities as an assessment tool and as an instructional strategy},
  journal = {Research and Practice in Technology Enhanced Learning},
  year    = {2015},
  volume  = {10},
  number  = {1},
  pages   = {5},
  issn    = {1793-7078},
  doi     = {10.1007/s41039-015-0006-0},
  url     = {https://doi.org/10.1007/s41039-015-0006-0}}

@inproceedings{yang2025learning,
title={Learning by Teaching: Engaging Students as Instructors of Large Language Models in Computer Science Education},
author={Xinming Yang and Haasil Pujara and Jun Li},
booktitle={Second Conference on Language Modeling},
year={2025},
doi={https://doi.org/10.48550/arXiv.2508.05979}
}

@article{Yan2025LLMcollaborative,
  author  = {Yan, Yi-Miao and Chen, Chuang-Qi and Hu, Yang-Bang and Ye, Xin-Dong},
  title   = {LLM-based collaborative programming: impact on students' computational thinking and self-efficacy},
  journal = {Humanities and Social Sciences Communications},
  year    = {2025},
  volume  = {12},
  number  = {1},
  pages   = {149},
  issn    = {2662-9992},
  doi     = {10.1057/s41599-025-04471-1},
  url     = {https://doi.org/10.1057/s41599-025-04471-1}}

@inproceedings{Logacheva2024ContextPersonalization,
author = {Logacheva, Evanfiya and Hellas, Arto and Prather, James and Sarsa, Sami and Leinonen, Juho},
title = {Evaluating Contextually Personalized Programming Exercises Created with Generative AI},
year = {2024},
isbn = {9798400704758},
publisher = {Association for Computing Machinery},
address = {New York, NY, USA},
url = {https://doi.org/10.1145/3632620.3671103},
doi = {10.1145/3632620.3671103},
booktitle = {Proceedings of the 2024 ACM Conference on International Computing Education Research - Volume 1},
pages = {95–113},
numpages = {19},
location = {Melbourne, VIC, Australia},
series = {ICER '24}
}

@article{Kim2026AIM,
  author  = {Kim, Young Rae and Park, Mi Sun and Joung, Eunmi},
  title   = {Exploring the integration of artificial intelligence in math education: Preservice Teachers' experiences and reflections on problem-posing activities with ChatGPT},
  journal = {School Science and Mathematics},
  year    = {2026},
  volume  = {126},
  number  = {1},
  pages   = {9--23},
  issn    = {0036-6803},
  doi     = {10.1111/ssm.18336},
  url     = {https://doi.org/10.1111/ssm.18336}
}

@article{KoedingerAleven2007Assistance,
  author={Koedinger, Kenneth R. and Aleven, Vincent},
  title={Exploring the Assistance Dilemma in Experiments with Cognitive Tutors},
  journal={Educational Psychology Review}, year={2007}, volume={19}, number={3},
  pages={239--264}, doi={10.1007/s10648-007-9049-0}
}

@article{vanDePol2010Scaffolding,
  author={van de Pol, Janneke and Volman, Monique and Beishuizen, Jos},
  title={Scaffolding in Teacher--Student Interaction: A Decade of Research},
  journal={Educational Psychology Review}, year={2010}, volume={22}, number={3},
  pages={271--296}, doi={10.1007/s10648-010-9127-6}
}

@article{SoderstromBjork2015LearningPerformance,
  author={Soderstrom, Nicholas C. and Bjork, Robert A.},
  title={Learning Versus Performance: An Integrative Review},
  journal={Perspectives on Psychological Science}, year={2015}, volume={10}, number={2},
  pages={176--199}, doi={10.1177/1745691615569000}
}

@article{AtkinsonRenklMerrill2003Fading,
  author={Atkinson, Robert K. and Renkl, Alexander and Merrill, Mary Margaret},
  title={Transitioning From Studying Examples to Solving Problems: Effects of Self-Explanation Prompts and Fading Worked-Out Steps},
  journal={Journal of Educational Psychology}, year={2003}, volume={95}, number={4},
  pages={774--783}, doi={10.1037/0022-0663.95.4.774}
}

@article{RenklAtkinson2003Transition,
  author={Renkl, Alexander and Atkinson, Robert K.},
  title={Structuring the Transition From Example Study to Problem Solving in Cognitive Skill Acquisition: A Cognitive Load Perspective},
  journal={Educational Psychologist}, year={2003}, volume={38}, number={1},
  pages={15--22}, doi={10.1207/S15326985EP3801_3}
}
\end{document}